\documentclass[showpacs,preprintnumbers,amsmath,amssymb,twocolumn,nofootinbib]{revtex4}
\usepackage{graphicx}% Include figure files
\usepackage{dcolumn}% Align table columns on decimal point
\usepackage{bm}% bold math
\usepackage{hyperref}
\usepackage{epsfig}
\newcommand{\nco}{\newcommand}
\nco{\beq}{\begin{equation}} \nco{\eeq}{\end{equation}}
\nco{\beqa}{\begin{eqnarray}} \nco{\eeqa}{\end{eqnarray}}
\nco{\lra}{\leftrightarrow}

\nco{\sss}{\scriptscriptstyle} \nco{\dphi}{\varphi}
\nco{\lsim}{\mbox{\raisebox{-.6ex}{~$\stackrel{<}{\sim}$~}}}
\nco{\gsim}{\mbox{\raisebox{-.6ex}{~$\stackrel{>}{\sim}$~}}}
\nco{\etal}{\textit{et al.}}
\nco{\ud}{\mathrm{d}}
\def\VEV#1{{\langle #1 \rangle}}

\begin{document}

%\preprint{McGill 03-25}

\title{Exciting dark matter in the galactic center}

\author{Fang Chen, James M.\ Cline,  Anthony Fradette, 
Andrew R.\ Frey, Charles Rabideau}

\affiliation{%
\centerline{Physics Department, McGill University,
3600 University Street, Montr\'eal, Qu\'ebec, Canada H3A 2T8}
e-mail: fangchen, jcline, frey\ @physics.mcgill.ca }

\date{November 11, 2009}

\begin{abstract}

We reconsider the proposal of excited dark matter (DM) as an
explanation for excess 511 keV gamma rays from positrons in the
galactic center. We quantitatively compute the cross section for DM
annihilation to nearby excited states, mediated by exchange of a new
light gauge boson with off-diagonal couplings to the DM states. 
In models where both excited states must be heavy enough to decay
into $e^+ e^-$ and the ground state, 
the predicted rate of positron production is never large 
enough to agree with observations, unless one makes extreme
assumptions about the local circular velocity in the Milky Way, or 
alternatively if there exists a metastable population of DM states
which can be excited through a mass gap of less than 650 keV, before
decaying into electrons and positrons.\\
 
\centerline{Dedicated to the memory of Lev Kofman}
\end{abstract}

% PACS, the Physics and Astronomy
                             % Classification Scheme.
%\keywords{Suggested keywords}%Use showkeys class option if keyword
                              %display desired
\maketitle

\section{Introduction}

There is presently much discussion about whether several anomalous
observations in galactic gamma ray and cosmic ray astronomy are better
explained by dark matter models or by more conventional astrophysics.
One such example is the 511 keV gamma ray excess
from the galactic center, which has been observed for 40 years, most
recently by the SPI spectrometer on the INTEGRAL satellite
\cite{spi}; for a recent review see ref.\ \cite{Diehl}.  Pulsars
\cite{pulsars}, gamma ray bursts \cite{grb},
supernovae \cite{sne}, low-mass x-ray binaries \cite{lmxrb},
the galactic black hole \cite{bh} and anisotropic propagation
effects \cite{prop}
 have been suggested as sources
of positrons whose annihilation could explain the observations, but
there is no consensus within the astrophysical community as to which
of these might be the right explanation.

There have also been many attempts to explain the 511 keV signal using
particle physics models, including annihilations of MeV scale dark
matter (DM) \cite{mevdm} or millicharged DM \cite{milli}, emission from cosmic strings
\cite{cosmic-strings}, decays of sterile neutrinos \cite{pp}, axinos \cite{axinos}, 
moduli (or modulinos)
\cite{moduli}, WIMPs \cite{PR,wimps}, light photinos \cite{photini},
or emission from composite objects \cite{composite}.
Excited dark matter (XDM)
\cite{xdm} is a particularly appealing example, in which heavy DM
particles in their ground state scatter into excited states,
$\chi_0\chi_0\to \chi_1\chi_1$, with mass difference $\delta M =
M_1-M_0$.  The excited states subsequently decay into $e^+
e^-\chi_0$, where the leptons are approximately nonrelativistic due
to the mass difference being close to the threshold value $\Delta
M\gsim 2 m_e$.  If the different DM states are members of a multiplet
of a new hidden gauge symmetry, this framework has the advantage of
being able to explain the small mass splittings naturally through
quantum loop effects \cite{nima}; moreover the same class of theories
can potentially explain excess positrons seen by higher energy
experiments including ATIC \cite{atic}, PPB-BETS \cite{ppb-bets},
PAMELA \cite{pamela} and the Fermi Large Area Telescope \cite{fermi}.

The XDM proposal requires that the excitation cross section be large,
in fact close to the unitarity bound, in at least a few (and possibly
many) partial waves
\cite{PR}, \cite{FSWY}.  Ref.\ \cite{xdm} made a first attempt to
achieve such large values in a model where the excitation was
mediated by exchange of a light scalar $\phi$ with $m_\phi$ in the
MeV$-$GeV range.  It was recognized that one must resum ladder
diagrams with multiple $\phi$ exchanges when the DM particles are
scattering at low velocity, but a quantitatively reliable way of
doing so was not yet appreciated at the time of this work. 

Subsequently ref.\ \cite{nima} showed how the calculation can be set
up in the framework of nonrelativistic quantum mechanics of a
two-state system.  It provides an example of the Sommerfeld
enhancement \cite{Sommerfeld,Slatyer}, which in the last few years has been widely studied in
the context of galactic dark matter annihilations.  However, unlike
the simpler one-state system where the enhancement can be
approximated analytically, in the two-state system no analytic
results for the excitation cross section are known (however see ref.\
\cite{Slatyer} for recent progress in the annihilation cross section
for multi-state DM).   Ref.\ \cite{CCF}  made a first attempt to
numerically solve the two-state system and to perform  a preliminary
scan of the parameter space.  Technical challenges limited that
analysis to relatively small coupling strengths of the exchanged
boson to the dark matter.  One goal in the present work is to extend
these results to stronger couplings and to explore more widely the
range of possibilities, including the dependence on the mass of the
exchanged boson.  A second is to  explore the dependence of the
predicted rate  on parameters of the dark matter density and velocity
profiles in the galaxy.  

In section \ref{setup} we review the nonrelativistic quantum
mechanical formulation of the problem and give a classical estimate
of the number of partial waves that can be expected to significantly
contribute to the excitation cross section.  In section \ref{method} we discuss
the numerical  method for computing partial wave amplitudes $f_l$ as
a function of velocity, and present sample results for $f_l$, as well
as a  survey of the range of boost factors which arise from a broad 
exploration of parameter space. In section \ref{rate_sect} we show how
these results go into the  calculation of the rate of positron
production in the galactic bulge; our choice of different DM density
and velocity dispersion profiles is explained there.  Section
\ref{survey} presents the  results  of our scan of the XDM model's
parameter space, consisting of the interaction strength and the mass
of the exchanged particle, and the  mass splitting between ground and
excited state.  We at first hold the DM mass $M_0$ fixed at a TeV
\cite{wimp-mass},
but then explore the dependence on  $M_0$ and galactic DM
distribution parameters at some optimal values of the microscopic
parameters.

%%**

Our results show that a large enough rate to match observations
cannot be achieved in the most straightforward implementation
of the XDM scenario, but in section \ref{alternative} we review
a modified version which can overcome this deficit.  In this version of XDM,
dark matter has at least three states; the middle state is stable or
metastable and can excite over a smaller gap to the highest state.
In section \ref{angular} we compare the predictions for the angular
distribution of 511 keV gamma rays with the observed signal.
Conclusions are given in section \ref{conclusion}.  Appendix \ref{born}
derives the Born approximation for the excitation cross section,
and \ref{vesc_app} reviews the derivation of the radial dependence of
the DM escape velocity.

\section{Excitation cross section}
\label{setup}

A key realization in ref.\ \cite{nima} is that the DM particles in
the galaxy are highly nonrelativistic, and so it is not necessary to
use the full apparatus of quantum field theory to analyze their
scattering.  Instead one can use the low energy effective theory,
which is quantum mechanics.  This enormously simplifies the problem,
getting around the need to resum a perturbative expansion in ladder
diagrams to find an effect which is nonperturbative, and enhanced by
$1/v$ (or $1/v^2$), the inverse velocity of the DM particles.  

\subsection{Quantum mechanical analysis}

We  assume that the DM is Majorana and that its
multiplicity arises from it being in a nontrivial representation of a
nonabelian gauge symmetry.  This immediately implies that any two
members of the multiplet, say $\chi_1$ and $\chi_2$, can only interact
via off-diagonal couplings to the gauge boson.  Furthermore the gauge
symmetry should be spontaneously broken so that $\chi_1$ and  
$\chi_2$ can be nondegenerate; then the exchanged vector boson gets 
a mass $\mu$.  This leads to an attractive Yukawa potential, but only
as an off-diagonal term in the matrix elements of the potential for
the two-state system.  
It is important to note that each state consists of two particles,
\beq
	|1\rangle = |\chi_1,\chi_1\rangle,\qquad
	|2\rangle = |\chi_2,\chi_2\rangle
\label{hilbert}
\eeq
by virtue of the fact that whenever a gauge boson is exchanged, the
color of both DM particles has to change.  There also exist the states
$|\chi_1,\chi_2\rangle$ and $|\chi_2,\chi_1\rangle$ which live in
their own superselection sector, 
but we do not
need to consider them because $\chi_2$ is presumed to not be present
in the initial state.   The diagonal terms have no interactions;
this part of the Hamiltonian consists only of the 
the mass-energies of the two particles.  Subtracting the mass
of the ground state, $2M_0$, the matrix potential in the basis (\ref{hilbert})
is
\beq
	V_{ij} = \left(\begin{array}{cc} 0 & -{\alpha_g e^{-\mu r}/r}\\
-{\alpha_g e^{-\mu r}/r} & 2\delta M \end{array}\right)
\label{Veq}
\eeq
where $\alpha_g=g^2/4\pi$ is the fine structure constant for the SU(N)
gauge coupling $g$. The wave function for the two-state system (with components labeled by
index $i$) in the CM frame
is $\Psi^i = \sum_{l}P_l(\cos\theta)R^i_{kl}(r)$, where $k$ is
the initial momentum.  Defining $\Phi_{l,i}(r) = R^i_{kl}/r$,
the Schr\"odinger equation is
\beq %%**
	-{1\over M_0}\Phi_{l,i}'' + \left({l(l+1)\over M_0 r^2}
	\delta_{ij}
	+ V_{ij}\right)\Phi_{l,j} = {k^2\over M_0} \Phi_{l,i}
\label{seq}
\eeq
For the numerical solution it is useful to rescale $r =
(\alpha_g/2\delta M) x$ and define the dimensionless variables 
\beq
\Gamma = M_0{\alpha_g^2\over 2\delta M},\quad
\Delta = {k^2\over 2 M_0\delta M} = {v^2\over v_t^2},\quad
\eta = {\alpha_g\mu\over 2\delta M}
\label{params}
\eeq
where $v_t=\sqrt{2\delta M/M_0}$ is the threshold velocity for producing the
excited state.  Then
the Schr\"odinger equation takes the form 
\beqa
	-\Phi'' &+&\left({l(l+1)\over x^2} + \Gamma(\hat V-\Delta)
	\right)\Phi = 0,\\
\label{numeq}
\hat V &=& \left(
	\begin{array}{cc} 0& -{e^{-\eta x}\over x}\\
	-{e^{-\eta x}\over x}& 1\end{array}\right)
\label{numpot}
\eeqa
with the dimensionless potential $\hat V$.  We require
$\Delta > 1$ for the initial state to have enough energy to
produce the heavier $|\chi_2,\chi_2\rangle$ final state.

%%**
To extract the scattering amplitudes, we decompose the numerical
solution into incoming and outgoing waves, $\Phi_{l,1}^{\rm in}$,
$\Phi_{l,1}^{\rm out}$ and $\Phi_{l,2}^{\rm out}$.  Partial wave
unitarity implies the conservation of flux, $k|\Phi_{l,1}^{\rm
in}|^2 = k|\Phi_{l,1}^{\rm out}|^2 + k'|\Phi_{l,2}^{\rm out}|^2$
(where $k'^2 \cong k^2 - 2 M_0\delta M$), which we use as a check on our
numerics.  The fraction of incoming $|\chi_1,\chi_1\rangle$ states
which gets converted to the $|\chi_2,\chi_2\rangle$ final state is
thus 
\beq
	f_l =  {k'\over k} {|\Phi_{l,2}^{\rm out}|^2\over 
	|\Phi_{l,1}^{\rm in}|^2}
\label{fl}
\eeq
in the $l$th partial wave.   $f_l$ provides a measure of the extent
to which a given partial wave can saturate the unitarity bound 
$f_l\le 1$ for
$\sigma_l$, its contribution to the cross section.   
The partial wave cross section is
\beq
	\sigma_l = {\pi(2l+1)\over M_0^2 v^2} f_l(v)
\label{stot}
\eeq
and the total cross section is $\sigma = \sum_l \sigma_l$.

\subsection{Classical treatment}
To get additional insight, it is useful to think about the classical
version of the problem, in the limit where the scattering
is elastic.  The standard method for solving the central
potential problem is to change variables to $u=1/r$ and to solve for
$\theta(u)$ instead of $r(t)$, where $\theta$ is the polar angle
in the scattering plane (see for example section 4.5 of \cite{hf}).  Take $\bar M = \frac12 M_0$ to be the
reduced mass.  Then it is straightforward to solve the first integral
of the motion for $\theta(u)$ with initial conditions where the
particles have velocity $v = v_{\rm rel}/2$ in the center-of-mass frame and impact
parameter $b$.  Taking $w = bu$,  the scattering angle is
\beq
	\theta_s = \pi - 2\int_0^{w_0} dw\left(1 - w^2 +
	{2\alpha_g\over\bar M b\, v_{\rm rel}^2} e^{-\mu b / w}\right)^{-1/2}
\label{theta_s}
\eeq
where $w_0\sim 1$ is the turning point of the effective potential, 
the point where the integrand diverges.  Notice that $\theta_s=0$
in the limit where the scattering potential vanishes.
The result (\ref{theta_s}) shows the classical origin of the
Sommerfeld enhancement, where the effect of the potential is
strengthened by the factor $1/v^2$ at low velocities, since the
particle has more time to be influenced when it is moving slowly.

We can estimate the maximum angular momentum $l=M_0 v b$ which 
gives significant
scattering by demanding that $(2\alpha_g/\bar M b v_{\rm rel}^2) e^{-\mu b}
\gsim 1$.  Solving for the argument of the exponential gives
\beq
	l_{\rm max} \sim {M_0\over\mu}v\, \ln\left(\alpha_g\over v 
	\,l_{\rm max} \right)\sim 
{M_0\over\mu}v\, \ln\left(\alpha_g\mu\over M_0 v^2 
	\, \right)
\label{lmax}
\eeq
The prefactor $M_0 v/\mu$ is what one would have obtained by
assuming the range of the force is $1/\mu$.  Substituting this
estimate for $l_{\rm max}$ (without the logarithm) into the formula for the cross section
(\ref{stot}) and assuming $f_l=1$ for all $l< l_{\rm max}$ results in
the geometrical value of the cross section, $\sigma = \pi/\mu^2$.  
The extra $\ln\left(\alpha_g\mu/ M_0 v^2 
	\, \right)$ factor is the origin of the Sommerfeld enhancement, which
boosts the cross section by the square of the logarithm for 
low-velocity scattering.

For the inelastic process, these classical insights need to be
modified, since there is a threshold $v> v_t$ for
production of the excited state.  Nevertheless they provide some 
idea as to how many partial waves one may expect to be important in
the quantum mechanical cross section (\ref{fl}).

\begin{figure*}[t]
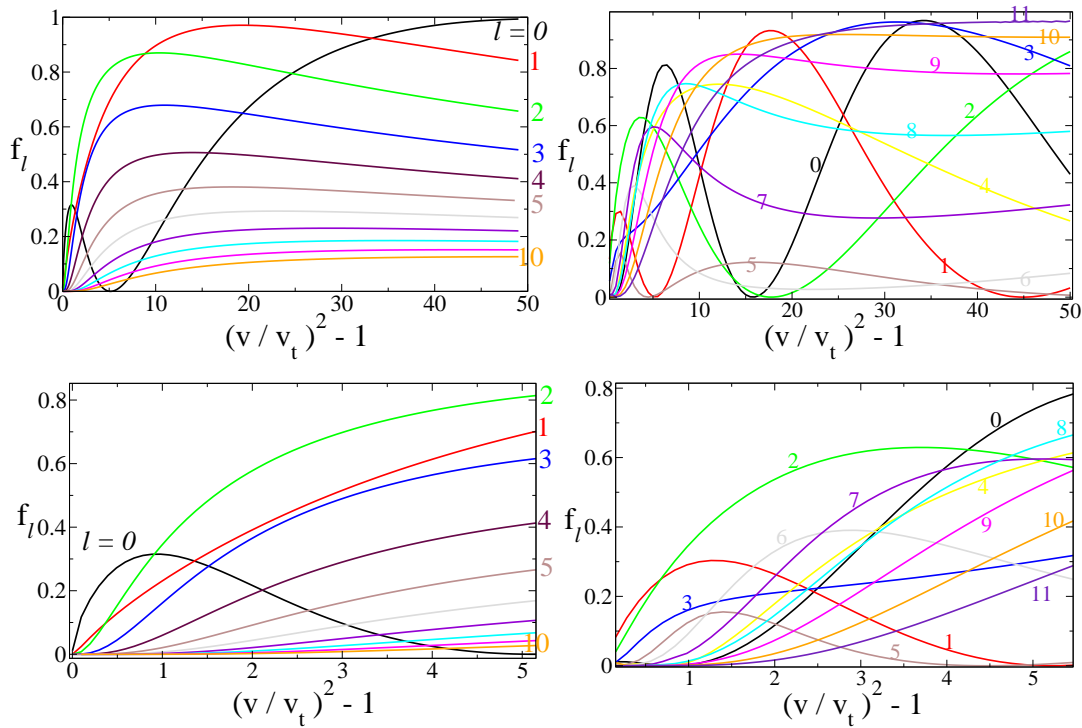

\smallskip \centerline{\epsfxsize=0.4\textwidth\epsfbox{res1.eps}
\epsfxsize=0.39\textwidth\epsfbox{res2.eps}}
\smallskip
\smallskip \centerline{\epsfxsize=0.4\textwidth\epsfbox{res1-exp.eps}
\epsfxsize=0.38\textwidth\epsfbox{res2-exp.eps}}
\caption{Top: fraction of partial waves converted to the excited state
as a function of center-of-mass velocity (relative to the threshold
velocity $v_t$) for two sample sets of parameters,
$\Gamma=10,\ \eta=1.2$, and $\Gamma=340,\ \eta=8.2$.  Each curve is
labeled by its partial wave $l$.  Bottom: same, but showing region 
of smaller velocities.}
\label{res}
\end{figure*}

\section{Partial wave amplitudes}
\label{method}

In this section we describe our method of numerical solution of 
the Schr\"odinger equation, leading to the partial wave amplitudes
$f_l(v)$.  
In each partial wave, one boundary condition is that there are only
incoming waves of state $|1\rangle$ from $r=\infty$, whereas the
outgoing waves are an admixture of $|1\rangle$ and $|2\rangle$.
Specifically, $|2\rangle$ must approach an eigenstate of radial momentum
with positive eigenvalue, 
\beq
 	-i{\partial\over \partial_r}\Phi_{l,2} = k'\Phi_{l,2}
\eeq
where $k' = \sqrt{k^2 - 2M_0\delta M}$.
This constitutes two
conditions, one for the real part and one for the imaginary part of
the wave function.   At the origin, both components behave like
$r^{l+1}$, but their relative amplitudes are not known, so one must
parametrize $\Phi_{l} \sim r^{l+1} ({1\atop b})$ with $b$ a complex
number.  We thus have a shooting problem: the real and imaginary
parts of $b$ must be adjusted so as to satisfy the two boundary
conditions at infinity.  Standard algorithms exist for this kind of 
problem; we adopt the routines in ref.\ \cite{numrec}.  

Instead of shooting, one can alternatively use a simpler approach,
which is to solve the Schr\"odinger equation with the correct
$r^{l+1}$ behavior near $r=0$, but arbitrary relative amplitudes of
$\Phi_{l,1}$ and $\Phi_{l,2}$, to obtain some solution $\Phi_0$,
which does not have the right behavior at large $r$.   Since the
complex conjugate $\Phi_0^*$ is also a solution, one can
algebraically construct the b linear combination $a\Phi_0+
b\Phi_0^*$ that has the desired behavior at large $r$.  This is much
faster than shooting because no iteration is required.

In principle, the method is straightforward, but complications arise
when one tries to consider parameters in the regime $\Gamma\gg 1$,
that is, $\alpha_g\gg \sqrt{2\delta M/M_0}$. Notice that this need not
be a particularly strong coupling for the application we have in
mind, where $\delta M\sim 2 m_e$ and $M\sim $ TeV: then $\alpha_g \gg
10^{-3}$.  The algorithms break down except when $\eta$, which
determines the dark gauge boson mass, is sufficiently large.   The
problem arises from the need to consider very different
scales  $1/\Gamma$ and $1/\eta$; in  particular the wave function
oscillates many times in the interaction region $x<1/\eta$.  Then (we
suspect) the solution with the desired behavior at large $r$ becomes
an exponentially small component of the generic solution, and so its
extraction gets lost in the numerical noise.  Despite this limitation,
we will be able to explore the parameter space widely enough to see
how the results extrapolate to the difficult regions, and thus 
give a complete characterization of the solutions.

%\begin{figure}[h]
%\smallskip \centerline{\epsfxsize=0.52\textwidth\epsfbox{res1-exp.eps}
%\epsfxsize=0.5\textwidth\epsfbox{res2-exp.eps}}
%\caption{Same as fig.\ \ref{res}, but showing region of smaller
%velocities.}
%\label{res-exp}
%\end{figure}

\begin{figure*}[t]
\smallskip \centerline{\epsfxsize=\textwidth\epsfbox{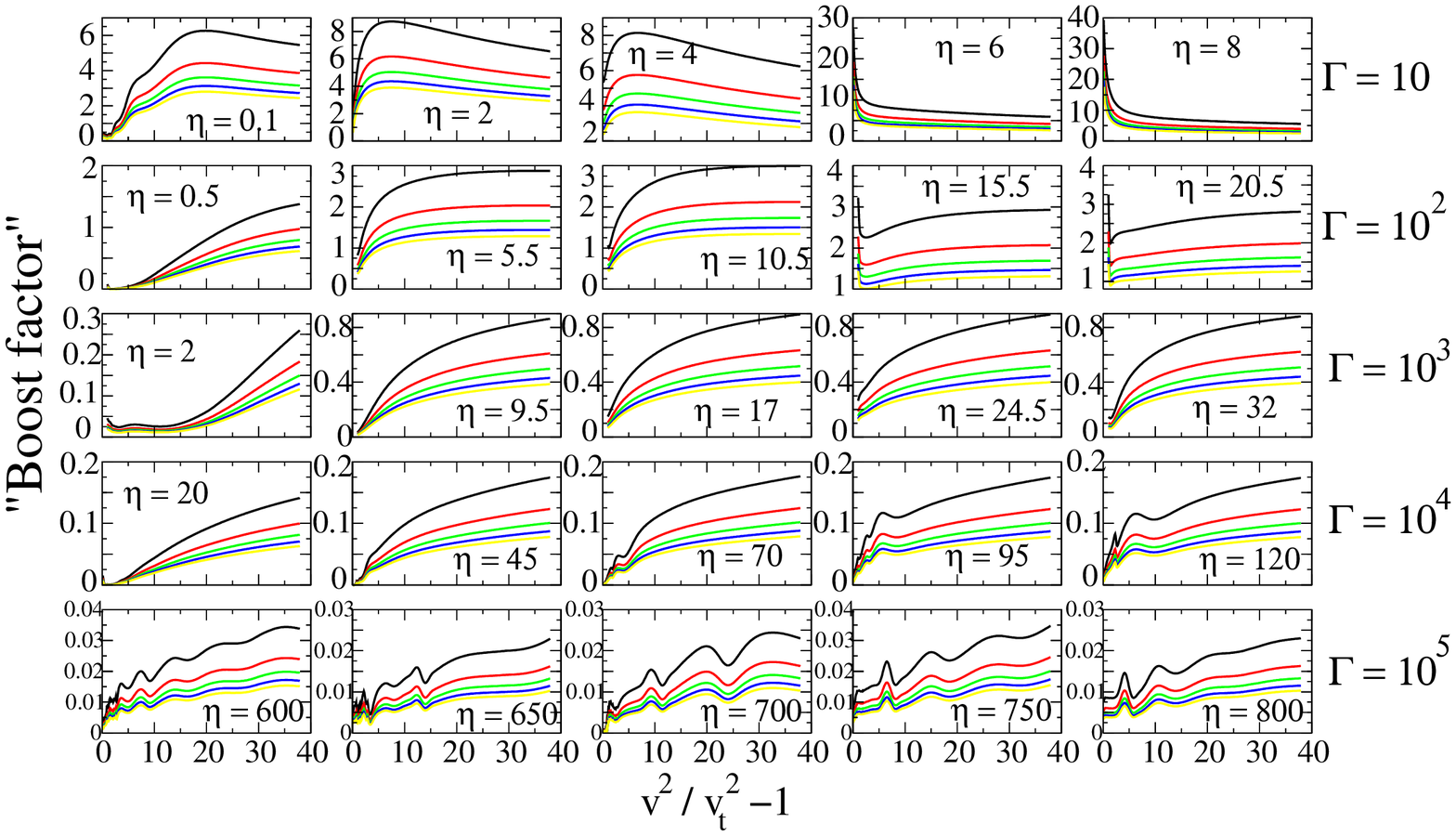}}
\caption{Boost factors as a function of the dimensionless squared
velocity $\Delta = v^2/v_t^2-1$ (recall that $v_t$ is the threshold
velocity) for a wide range of the 
parameters defined in eq.\ (\ref{params}),
 $\Gamma = 10,\dots ,10^5$ (rows) and $\eta\sim$ few
$\times \sqrt{\Gamma}$ (columns).  We assume $M_0 = 1$ TeV, and 
the curves in each panel correspond to $\delta M = 0.2,\dots, 1$ MeV,
from top to bottom.  
}
\label{boost}
\end{figure*}

\subsection{Results for $f_l$}

We show two examples of the results for $f_l$, eq.\ ({\ref{fl}) in
figure \ref{res}, as a function of $\Delta -1$ = $v^2/v_t^2 -1$;
recall that $v$ is the  velocity of one of the particles in the
center of mass frame, and $v_t$ is the threshold
velocity for producing the excited DM states.  These results look
quite different from the usual single-state $s$-wave Sommerfeld
enhancement, which has numerous sharp resonance bands at low 
velocity.  Here, there can be at best a few oscillations in a given
partial wave, but not strongly peaked.   These might be due to having
an approximately integral number of wavelengths within the width of
the potential well, which of course is not square and so one would
not expect nearly bound resonances to be very sharp.   It would be
interesting to find an analytic approximation to help better
understand these results.  We tried to develop the WKB
approximation in this context, but without 
success.\footnote{At least in some regions of parameter space, the WKB approximation fails
due to the breakdown of matching conditions at the classical turning point.  It is possible that more sophisticated
approximate analytical techniques such as those used in 
\cite{Slatyer} ameliorate this problem.}

For clarity in figure \ref{res} we have chosen examples where the
number of partial waves which contribute significantly are relatively
small, but for our actual computations, there are examples
(particularly for small values of $\eta \ll\sqrt{\Gamma}$) where
hundreds of partial waves are relevant.  Using the estimate
(\ref{lmax}), and the definitions (\ref{params}), we find
\beq
	l_{\rm max}\sim
{\sqrt{\Gamma\Delta}\over\eta}\ln{\eta\over\Delta}
\eeq
We typically computed up to
$l_{\rm max} = 500$ to insure convergence of the sum over $l$ in the
cross section, and we restrict ourselves to gauge boson masses, hence
values of $\eta\sim \sqrt{\Gamma}$, which are large enough to justify neglecting higher
values of $l$ (it also turns out that the interesting functional
dependence of $\sigma$ on $\eta$ occurs in the region where  $\eta\sim
\sqrt{\Gamma}$).   It will be apparent from the results that this is a
mild restriction, in terms of understanding trends across the full
parameter space of the model.

\subsection{Boost factors}

There is a large literature on the Sommerfeld enhancement for 
$s$-wave annihilation of DM at low velocities, where the effect is
characterized by a boost factor, defined as the ratio of the enhanced cross 
section to the Born approximation value.  Even though all our results
for the scattering cross section are computed directly, without
reference to the perturbative value, one might nevertheless be interested to
know their ratio  in the present context.  To make the comparison,
we need the perturbative expression for $\sigma$ (here computed using
relativistic quantum field theory),
\begin{eqnarray}
\sigma &=&\frac{\pi\alpha_g^2}{2\Delta} \Bigg\{ \frac{32}
{\left(M_0^2v_t^2(2\Delta-1)+\mu^2\right)v_t^2}
\ln{1+\epsilon\over 1-\epsilon}\nonumber\\
%&&\left.\ln\left(\frac{\mu^2+M_0^2v_t^2(2\Delta-1)
%+2M_0^2v_t^2\sqrt{\Delta(\Delta-1)}}
%{\mu^2+M_0^2v_t^2(2\Delta-1)-2M_0^2v_t^2\sqrt{\Delta(\Delta-1)}}
%\right)\right.
%\nonumber\\
&+&\frac{128M_0^2\sqrt{\Delta(\Delta-1)}}
{\mu^4+M_0^4v_t^4+2\mu^2M_0^2v_t^2(2\Delta-1)}\Bigg\}
\label{sigma_born}
\end{eqnarray}
where $\epsilon = 2M_0^2v_t^2\sqrt{\Delta(\Delta-1)} /
(\mu^2+M_0^2v_t^2(2\Delta-1))$, and 
we have assumed $\delta M\ll\mu\ll M_0$ to simplify the
expression; see appendix \ref{born} for details.  

In figure \ref{boost} we plot the resulting boost factors as a
function of center-of-mass velocity of the incoming particles,  for
the same range of  parameters that we explore in the next section as
being relevant  for positron production, namely $\Gamma =
10,\,\dots,10^5$ and $\eta\sim$ few$\times \sqrt{\Gamma}$, for $\delta
M = 0.2,\,0.4,\,\dots,\, 1$ MeV, and  assuming a DM mass of $M_0=1$
TeV.   From the figure it is apparent that the ``boost factor'' is an
enhancement only for the smaller values of $\Gamma$, and is actually
a suppression factor for $\Gamma>100$, in the range of velocities
where we have computed.  One can see the onset of resonant
enhancement at low velocity in the cases of $\Gamma = 10$ and 100,
where the boost factor suddenly increases as a function of the
dimensionless gauge boson mass parameter $\eta$.  For fixed 
parameters $\Gamma$, $\eta$, $\Delta$, we see from each family of
curves that smallest values of $\delta M$ give the largest cross
section.  This can be understood as being due to Sommerfeld enhancement,
since for fixed $\Delta$ decreasing $\delta M$ corresponds to
decreasing the particle velocities.

One check on these results is provided by considering when the 
Born approximation for our potential should be valid.  By demanding
that the scattered wave be small compared to the incoming wave
in nonrelativistic quantum mechanics, and working in the limit of
low velocity, one obtains the constraint 
$2M_0|\int_0^{\infty} r V(r) dr|\ll 1$, which implies $2\alpha_g M_0
\ll \mu$, or in terms of the parameters (\ref{params}), $2\Gamma \ll
\eta$.  This is violated everywhere in the region of parameter space
we have considered, so there is no contradiction between our results
and expectations based on the Born approximation.

\section{Rate of positron production}
\label{rate_sect}%%**
The rate of positron production within a radius $r_c$ of the 
galactic center is given by
\beq
R_{e^+} = \frac12 \int_0^{r_c}  
\VEV{\sigma v_{\rm rel}} n^2(r)\, 4\pi r^2 \,\ud r\,
\label{rate}
\eeq
where $n(r)= \rho(r)/M_0$ is the DM number density and $\rho(r)$ is
its mass density.  The factor of $\frac 12$ is to avoid
double-counting, and we have assumed that the DM is distributed with
spherical symmetry.  Since the INTEGRAL signal has a full-width
at half-maximum of 8$^\circ$ \cite{weid}, we take $r_c$ to be the radius which 
subtends 8$^\circ$ at our distance of $R_0 = 8$ kpc from the galactic
center, giving $r_c=1.1$ kpc.  We will discuss the uncertainty
in $R_0$ and its impact on our results below.  The predicted rate
(\ref{rate}) is to be compared to the observed one (see ref.\
\cite{xdm} for a detailed discussion) of $R_{\rm obs} = 3.4\times 10^{42}$
for the component coming from the galactic bulge \cite{xdm}.

The cross section in (\ref{rate}) is averaged over the
velocities of the DM particles, using the Maxwell-Boltzmann
distribution function
\beq 
f(v,r)= N\left\{ \begin{array}{ll}
v^2\exp\left(-{v^2}/{2\sigma_v(r)^2}\right), & v < v_{\rm esc}\\
0, & v \geq v_{\rm esc},
\end{array} \right.
\label{mbdist}
\eeq
with a cutoff at the escape velocity for galactic DM.  The
normalization factor is given by
\beq
	N^{-1}(r) = 2^{5/2}\pi \sigma_v^3(r)\left({\sqrt{\pi}\over 2}
	{\rm erf}(X) - Xe^{-X^2}\right)
\eeq
where $X = v_{\rm esc}/(\sqrt{2} \sigma_v(r))$.
We follow
ref.\ \cite{xdm} in using the $r$-dependent escape velocity
(see appendix \ref{vesc_app} for derivation),
\beqa
v_{\rm esc} &=& \left\{ \begin{array}{ll}
v_c\sqrt{2\left[1-\ln\left(\frac{r}{r_{-2}}\right)\right]}, &
                        r \leq r_{-2}\\
                        v_c\sqrt{2\frac{r_{-2}}{r}}, &
                        r > r_{-2}.
                   \end{array} \right.
\label{vesc_eq}
\eeqa 
$v_c$ is the circular velocity of DM, assumed to be approximately
constant at radii near the characteristic scale $r_{-2}$, 
which is defined below.   We take the 
fiducial value for $v_c$, 
\beq
	v_c = 220 \, {\rm km/s}
\eeq
as in ref.\ \cite{xdm} and many other references, but below 
we will also explore the sensitivity of our
predictions to changes in this value.  To average over the
cross section, one integrates over both DM particle velocities,
\beqa
\VEV{\sigma v_{\rm rel}}(r) &=&  
8\pi^2 \int_0^{\infty}  \ud v_1 \int_0^{\infty}  \ud v_2 
\int_{-1}^{1} \ud(\cos\theta) \nonumber\\
&&f(v_1,r)\, f(v_2,r)\, \sigma v_{\rm rel},
\label{eq_sigvrel}
\eeqa
where $v_{\rm rel} = (v_1^2 + v_2^2 - 2 v_1 v_2\cos\theta )^{1/2}$,
and the cross section is taken to vanish if $\frac12 v_{\rm rel}$
is less than the threshold velocity defined in eq.\ (\ref{params}).
We perform the integrals numerically.

\subsection{Pure DM radial distributions}

\begin{figure}[t]
\smallskip \centerline{\epsfxsize=0.5\textwidth\epsfbox{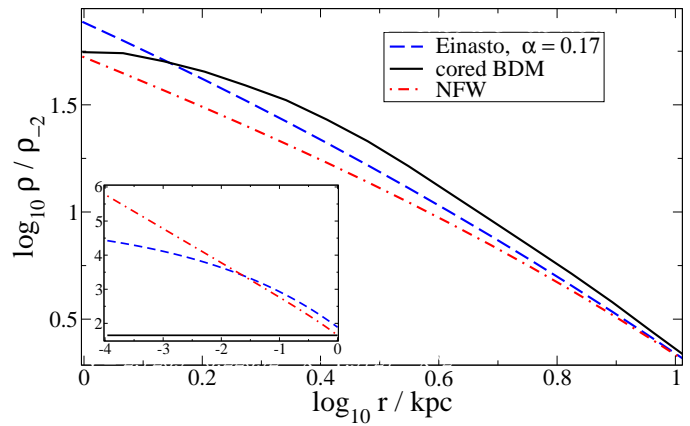}}
\caption{NFW, Einasto, and cored BDM density profiles, in the inner
$1-10$ kpc.  Inset shows the region $0.01-1$ kpc.}
\label{rhos}
\end{figure}

The $r$ dependence of the velocity dispersion $\sigma_v(r)$ and
also the escape velocity $v_{\rm esc}(r)$ appearing in (\ref{mbdist})
depend on the shape of the DM density profile $\rho(r)$.  We will
consider several hypotheses for the form of $\rho(r)$, inferred from
$N$-body simulations, some based on pure DM (PDM), and
others which contain a baryonic component in addition to the DM (BDM).   
The PDM Aquarius simulation \cite{Navarro}
finds the relation $\sigma_v^3 \propto r^{1.875} \rho(r)$, where $\rho$
is the DM density.  Thus one needs to specify the density profile
in order to fix $\sigma_v(r)$.

We consider two widely-used density profiles, the NFW form
\beq
\rho(r) = {4 \rho_{-2}}
{\left(\frac{r}{r_{-2}}\right)^{-1}\left(1+\frac{r}{r_{-2}}\right)^{-2}}\;,
\label{nfw}
\eeq
and the Einasto form
\beq
\rho(r) = \rho_{-2} \,
\exp{\left[\frac{-2}{\alpha}\left(\left(\frac{r}{r_{-2}}\right)^{\alpha}-1\right)\right]},
\label{einasto}
\eeq
Here $r_{-2}$ is the
radius at which the logarithmic slope of the density is $-2$, the
normalization is defined by
$\rho_{-2}=\rho(r_{-2})$, and  $\alpha$ is the shape
parameter.  A fit of the NFW form to the Milky Way galaxy
in ref.\ \cite{Battaglia} gives\footnote{In the notation of
ref.\ \cite{Battaglia}, $r_{-2} = r_s = r_v/c$ and $4\rho_{-2} = 
\delta_c \rho_c$, where $r_v$
is the virial velocity, $c$ is the concentration parameter,
$\delta_c = 100 c^3/[3(\ln(1+c) - c/(1+c)]$ and 
$\rho_c$ is the present critical density of the universe.
$r_v$ is related to the virial mass $M_v$ by
\[	r_v = \left({3M_v\over 4\pi\cdot 100\rho_c}\right)^{1/3}.
\]
Using their best fit values $c=18$, $M_v = 9.4\times 10^{11}M_\odot$
(see erratum of \cite{Battaglia}) leads to the values quoted for
$r_s$ and $\rho_{-2}$.}
 $r_{-2} = 14.1\, (h/0.7)^{-2/3}$ kpc
and $\rho_{-2} = 0.13\,(h/0.7)^2$ GeV/cm$^3$ where $h$ is the 
Hubble parameter.  These are close to the
best-fit values from the Aquarius
simulation for the Aq-A-1 galaxy, which was the highest resolution
simulation considered, given in tables 1-2 of ref.\ \cite{Navarro},
\beqa
	r_{-2} &=& 15\,{\rm\ kpc}\left({h\over 0.7}\right)^{-1},\quad 
	\rho_{-2} = 0.14\,{{\rm GeV}\over
	{\rm cm}^3 c^2}\left({h\over 0.7}\right)^2,
	\nonumber\\ \alpha &=& 0.17\qquad\qquad \hbox{(Aquarius-A-1
parameters})
\label{aqa1}
\eeqa
We  adopt these as our fiducial values.  The resulting
velocity dispersion is then
\beq
	\sigma_v = v_0 \left({r\over r_{-2}}\right)^{0.625}
	\left({\rho(r)\over\rho_{-2}}\right)^{1/3}
\label{vpdm}
\eeq
where the velocity scale $v_0$ (called $\sigma_{\rm max}$ in Table
1 of ref.\ \cite{Navarro}) is determined to be
\beq
	v_0 \cong 260\, {\rm km/s}
\label{v0eq}
\eeq
The radial dependence of $\sigma_v$ is plotted in figure \ref{sigv},
showing that the Einasto profile leads to much higher velocities
in the inner regions $r < r_{-2}$ than does the NFW profile.  

\begin{figure}[t]
\smallskip \centerline{\epsfxsize=0.44\textwidth\epsfbox{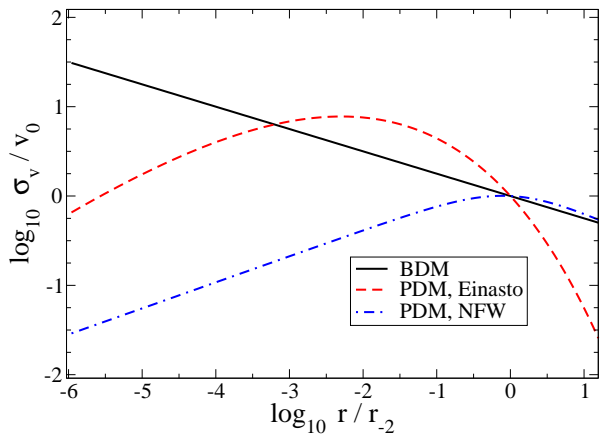}}
\caption{Velocity dispersion profile for dark matter $+$ baryons
(BDM), solid curve, and for pure dark matter with Einasto or NFW density
profiles, dashed curves.}
\label{sigv}
\end{figure}

\begin{figure*}[t]
\centerline{\epsfxsize=0.55\textwidth\epsfbox{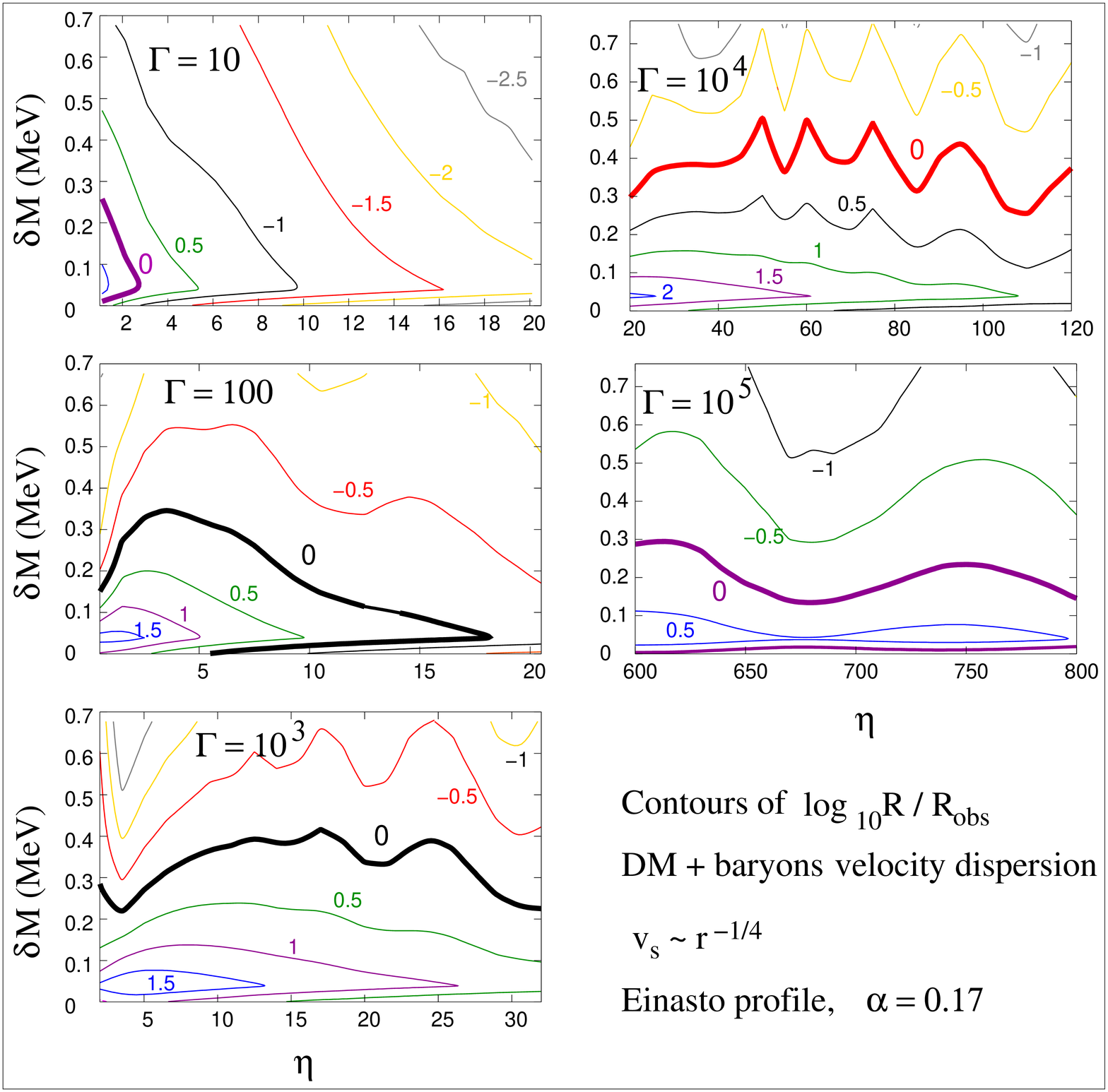}
\epsfxsize=0.55\textwidth\epsfbox{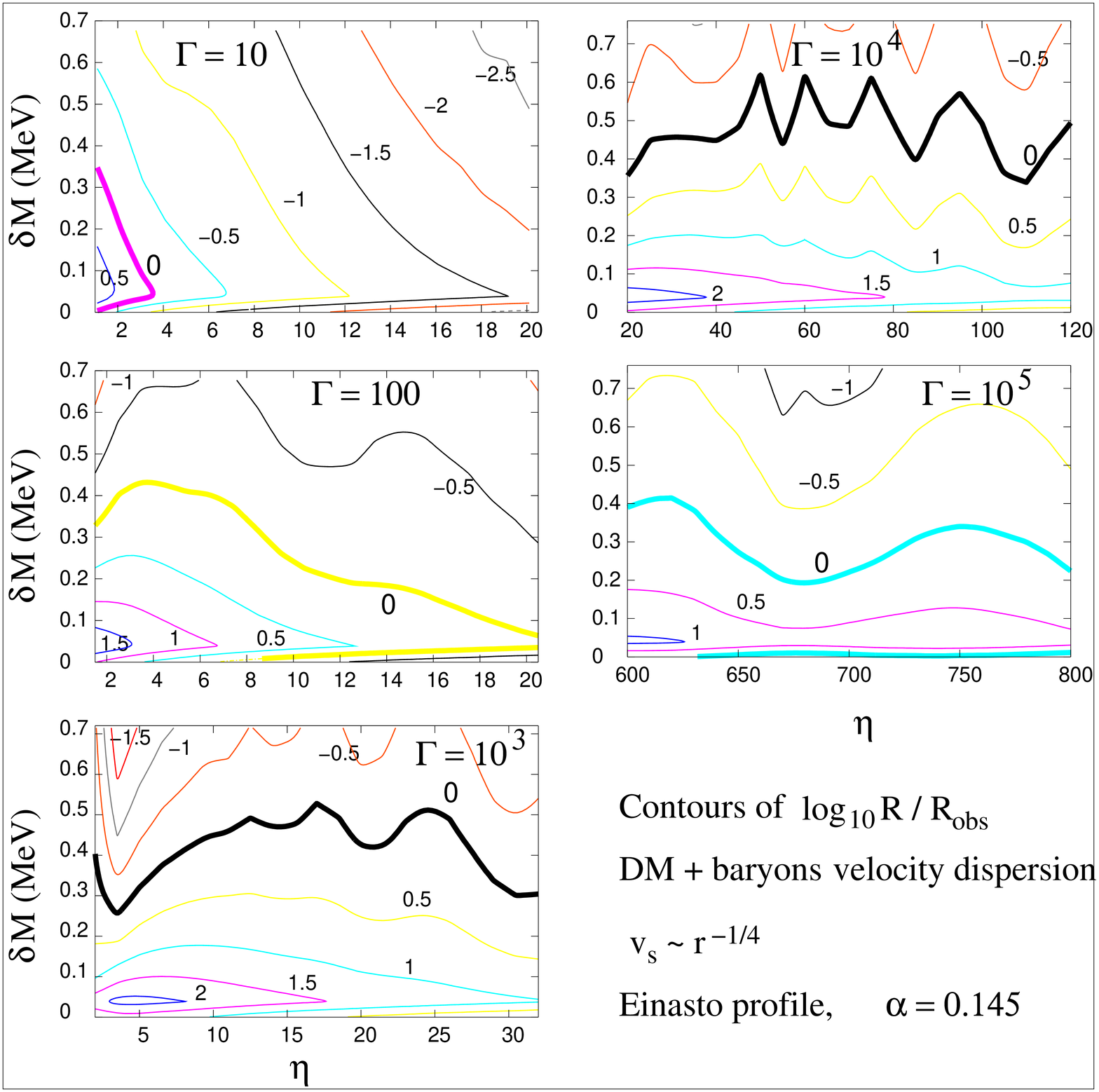}}
\centerline{\epsfxsize=0.55\textwidth\epsfbox{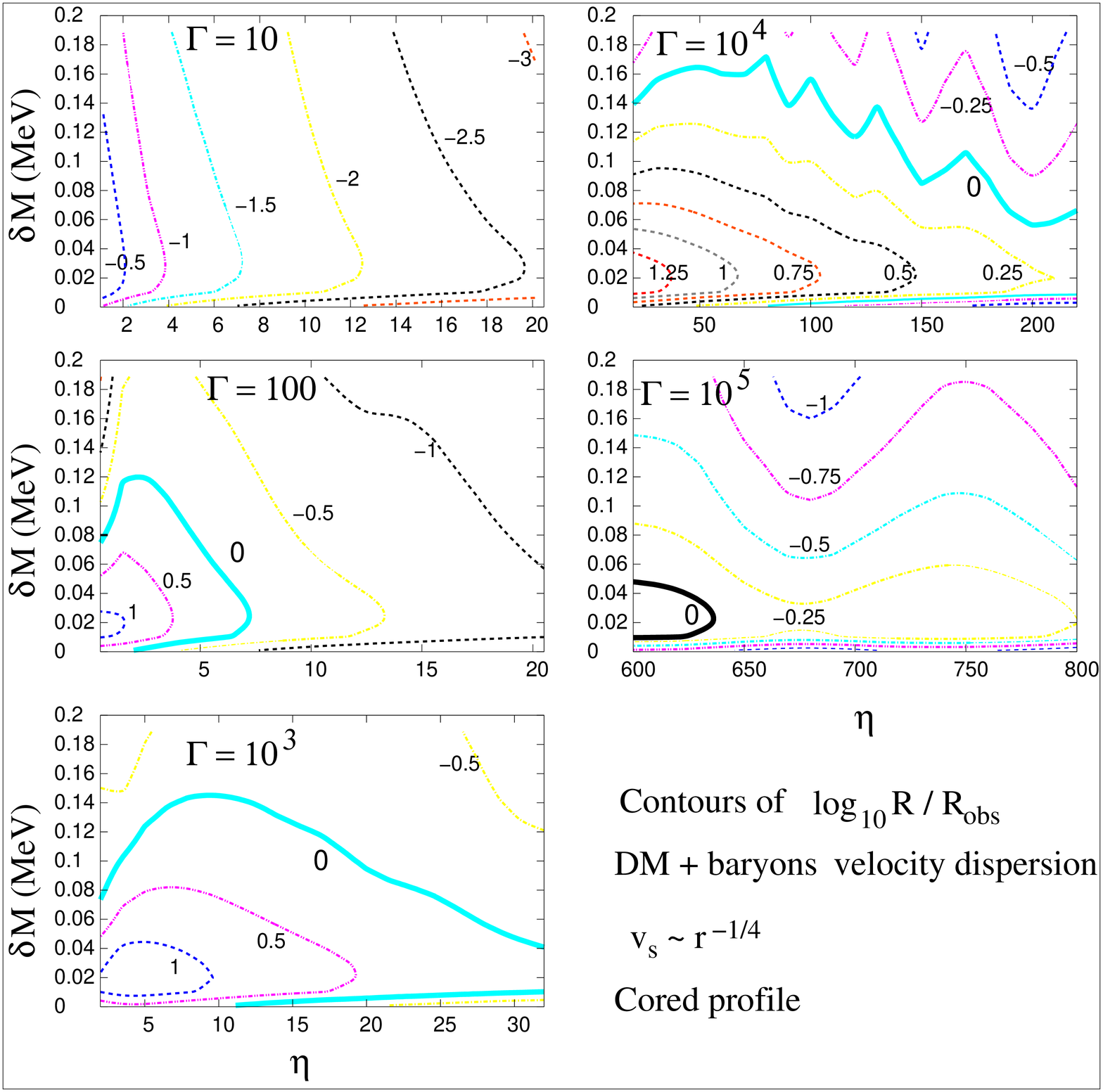}
\epsfxsize=0.55\textwidth\epsfbox{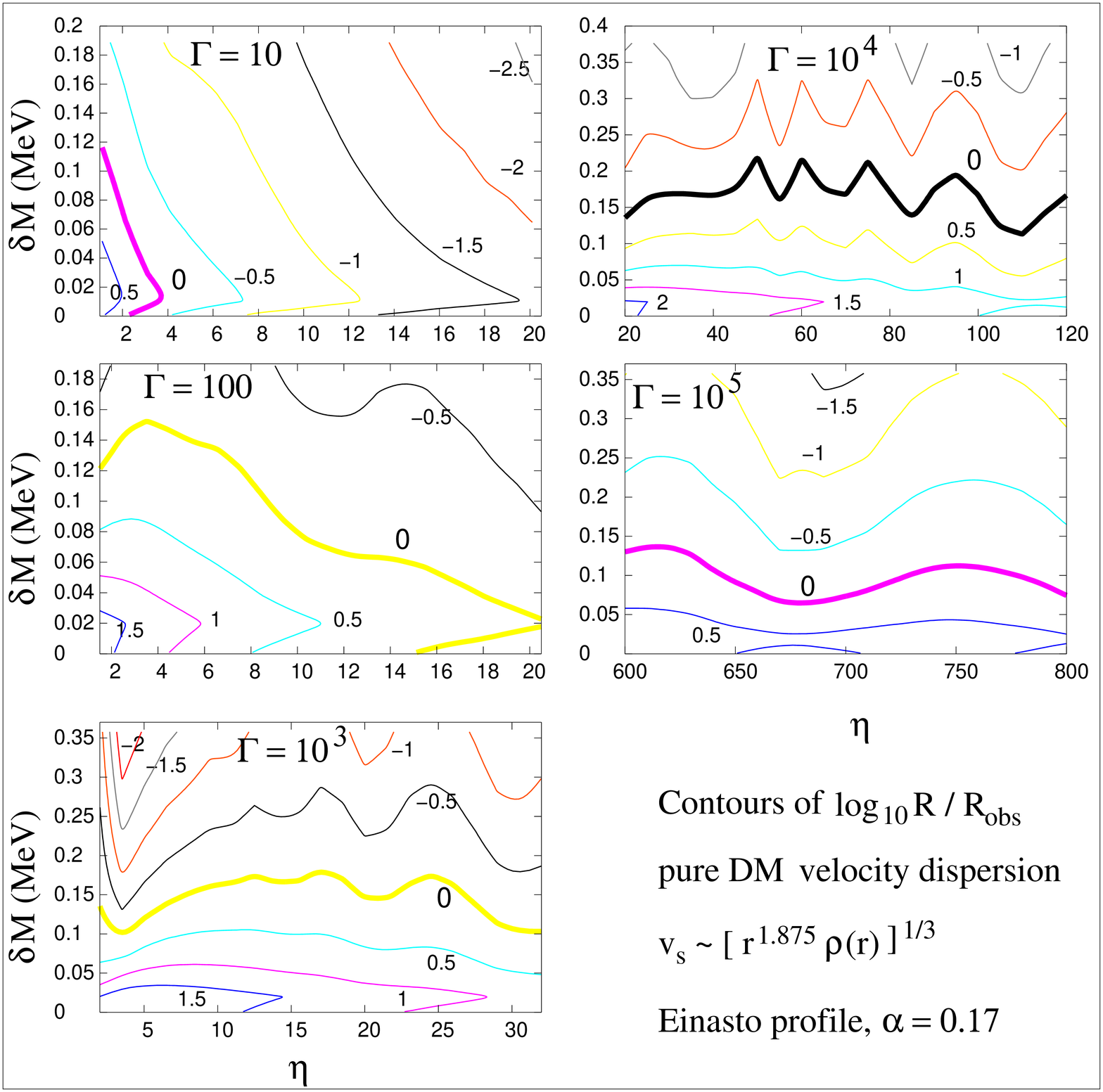}}
\caption{Contours of logarithm of predicted over observed rate of positron
production in the plane of $\eta$-$\delta M$ for several DM velocity
and density profiles.  ($\eta$ is proportional to the exchanged
gauge boson mass; see eq.\ (\ref{params}).)  
 Those labeled ``0'' (thickest contours) match the observed rate.
The DM mass is assumed to be $M_0 = $ 1 TeV.}
\label{edmcont}
\end{figure*}

\subsection{Effect of baryons on distributions}

The Aquarius simulation ignored the effects of baryons on the
evolution of galaxies, but recent
studies have included them and shown that their effect is to increase
the velocity dispersion in the inner region $r \lsim
10$ kpc
\cite{Romano}-\cite{Pedrosa}.  Fitting the results of figure 2 in
 ref.\ \cite{Romano}, we find the velocity dispersion
\beq
	\sigma_v(r) = v_0 \left({r\over r_{-2}}\right)^{-1/4}
\label{vbdm}
\eeq
where $v_0$ happens to take the same value as in (\ref{v0eq}).
Its shape is plotted in figure \ref{sigv}.  
The rise in $v$ toward the galactic center can greatly boost the
excitation rate of DM relative to the pure dark matter case,
(\ref{vpdm}), and we will employ both for the purposes of comparison.

There is some disagreement between the different BDM
simulations as to the impact of the baryons on the DM density
profile.  Ref.\ \cite{Romano} finds that the cusp of the density
distribution is softened (we refer to it as a ``cored'' profile), while ref.\ \cite{Pedrosa} obtains 
profiles that are consistent with the Einasto form.  We will consider
both possibilities in our computations of the positron production
rate.  To study the possible softening effect, we have digitized the
BDM density profile at redshift $z=0$ in fig.\ 1 of ref.\ \cite{Romano}
and fit its logarithm to a quartic polynomial,
\beq
	\log_{10}\rho_{\sss cored} = \sum_{n=0}^4 a_n \log_{10}^n(r/{\rm kpc})
\eeq
with coefficients $a_n = 1.7487,$ $0.0395,$ $-2.537$, $1.459$,
$-0.3448$, respectively, for $n=0,\dots,4$.  This profile is plotted
along with the NFW and Einasto forms (\ref{nfw})-(\ref{einasto})
in figure \ref{rhos} in the region 1 kpc $< r < 10$ kpc.  For larger
$r$, the three shapes are in closer agreement, while for $r < 1$ kpc
the BDM profile is taken to be constant.

Other BDM simulations do not find the erasure of the cusp; ref.\ 
\cite{Pedrosa} finds good fits to the Einasto profile with a range
of $\alpha$ values that are consistent with (\ref{aqa1}), but that
go as low as $\alpha=0.145$.  To illustrate the effect of the $\alpha$
parameter on the rate, we will consider this lower value in addition to
the higher one $\alpha=0.17$ in (\ref{aqa1}).

\subsection{Consistency of velocity distributions}
%%**

Strictly speaking, the velocity distributions (\ref{vpdm},\ref{vbdm}) 
may not be self-consistent
when varying the dark matter density profile.  A better approximation to 
the self-consistent distribution could be derived from the Jeans 
equation, following the technique of \cite{rz}, for example.  However,
our results are much more sensitive to the shape of the density profile
than the detailed velocity profile, so an improved treatment of the
velocity profile should only lead to qualitatively small changes 
in our results.
Figure \ref{others}, for example, shows that the rate of positron production
is relatively insensitive to the scale of the BDM velocity distribution
$v_0$ near or above the standard value (\ref{v0eq}).

We do however, find an enhancement in positron production in going from the
PDM velocity dispersion profile (\ref{vpdm}) to the more cuspy BDM profile 
(\ref{vbdm}).  We therefore might worry about overestimating the positron
production rate if a self-consistent treatment removes the cusp in the 
profile.  We believe it is more conservative to consider the cuspy profile
(\ref{vbdm}), as we are interested in determining whether or not the XDM 
mechanism is viable for $\delta M\gtrsim 2m_e$.  Even with the advantageous
profile (\ref{vbdm}), we still find that the positron production rate is only
sufficient for smaller $\delta M$.

\begin{figure*}[t]
\smallskip \centerline{\epsfxsize=0.7\textwidth\epsfbox{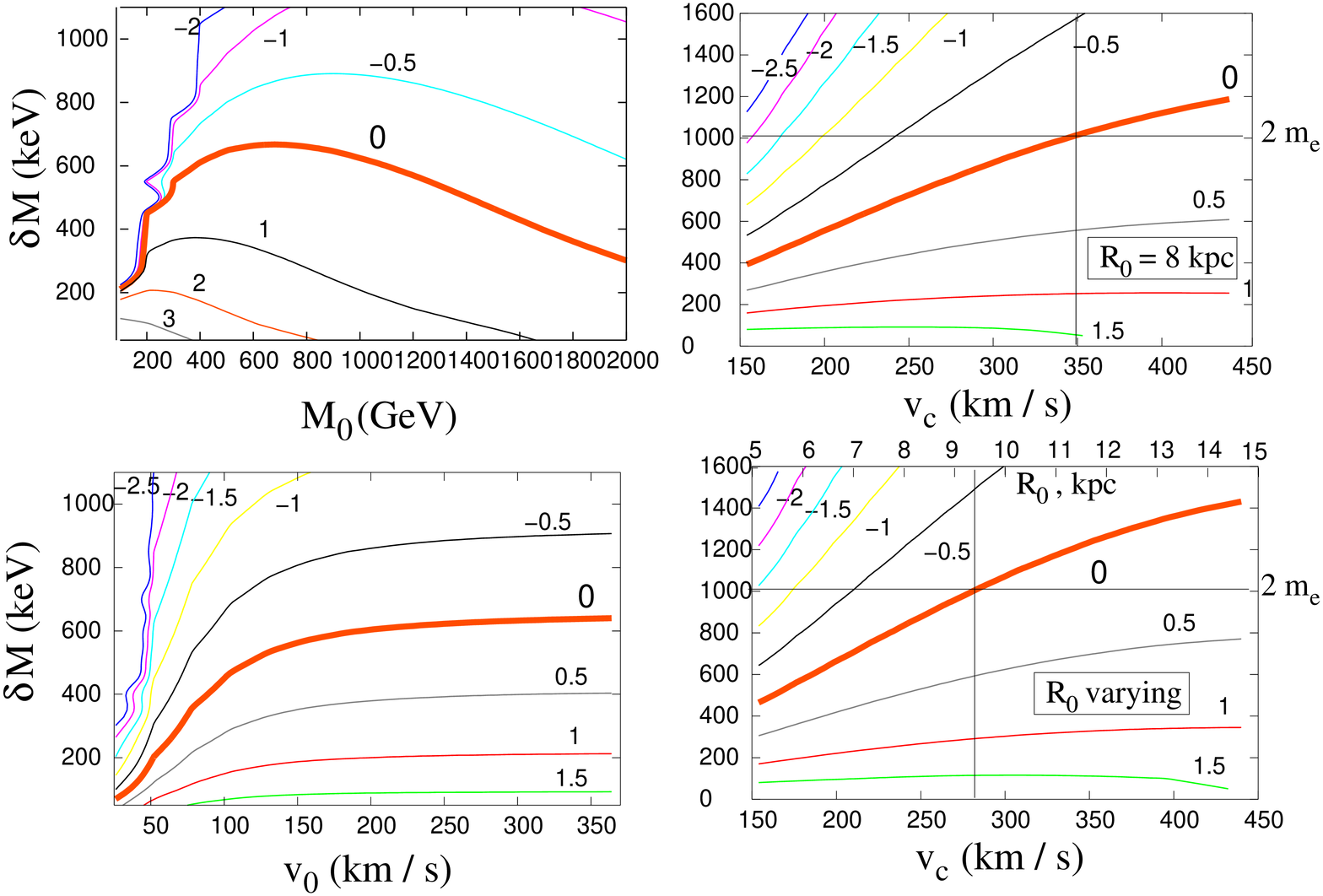}}
\caption{Log of predicted over observed rate contours in the plane of mass
splitting $\delta M$ versus $M_0$ (DM mass), $v_0$ (velocity
dispersion at $r_{-2} = 15$ kpc) 
or $v_c$ (circular velocity at $r_{-2}$). The contour labeled ``0'' 
corresponds to the observed rate of positron production.
The distance to the galactic
center is fixed at $R_0 =8$ kpc in the top right panel (and the left
ones), whereas it
varies linearly with $v_c$ in the bottom right one.  Other
parameters are fixed at $\Gamma=10^4$, $\eta=50$, $\alpha=0.145$ to 
maximize the rate, and the BDM velocity profile is used. }
\label{others}
\end{figure*}

\section{Survey of parameter space}
\label{survey}

There are four important dimensionless parameters that determine the
rate of scatterings to produce the XDM states in the galaxy. Two have
already been defined in eq.\ (\ref{params}),  $\Gamma$ and $\eta$,
which depend only upon the microphysics,  {\it i.e.,} the DM mass,
mass splitting, gauge boson mass, and interaction strength.  The
others depend on the ratio between the threshold velocity $v_t$ or
the escape velocity $v_{\rm esc}$ and the parameter $v_0$
(that controls the average speed of DM particles in the galaxy),
motivating us to define
\beq
    a \equiv {2\delta M\over M_0 v_0^2} = {v_t^2\over v_0^2},\qquad
	b = {v_{\rm esc}^2\over v_0^2}
\label{abdef}
\eeq
Dependence on $a,b$ arises from averaging the cross section over the
velocity distribution of the DM.  In ref.\ \cite{PR}, a simpler
estimate of the rate was made using the  isothermal  distribution function, 
$f(v) = N v^2 e^{-v^2/v_0^2}$, giving
\beq
	\langle \sigma v_{\rm rel} \rangle = 
 {2\sqrt{\pi}\over
M_0^2 v_0} \sum_l (2l+1) \int_{v_t^2}^{v_{\rm esc}^2} {dv^2\over
v_0^2}  
e^{-v^2/v_0^2} f_l(v)
\eeq
where $v_{\rm esc}$ is the escape velocity, also taken to be constant
and in the range $\sim 500-600$ km/s.\footnote{This estimate assumes
that $v_{\rm esc}\gg v_0$ so that $N\sim v_0^{-3}$.  
In the unphysical limit $v_{\rm esc}\ll v_0$,
the normalization factor goes to $N\sim v_{\rm esc}^{-3}$.
\label{f2}}
In the limit of large $a,b$, the rate is
exponentially suppressed, $\sim (e^{-a}-e^{-b})$.  Although we are
calculating the rate more quantitatively  in this paper, the simpler
approach makes it clear how to expect the results to depend on $a,b$,
at least semiquantitatively.  In terms of dimensionful constants and the
dimensionless ratios, one can parametrize the rate in the form
\beq
	R_{e^+} = {\rho_{-2}^2 r_{-2}^3\over M_0^4 v_0}\, 
	g(\Gamma,\eta,a,b)
\label{analytic}
\eeq
where the function $g$ of the dimensionless variables has complicated
dependence on $\Gamma$ and $\eta$, but roughly $(e^{-a}-e^{-b})$ dependence
on $a,b$.   The contours of $\log(e^{-a}-e^{-b})$ in the $a$-$b$ plane
are roughly linear over some ranges, suggesting that there should be
a quasilinear degeneracy between the parameters $\delta M$ and $v_{\rm esc}^2$,
which we will observe below.

\subsection{Dependence on $\Gamma$, $\eta$, $\delta M$}

For our initial exploration of parameter space, we  computed the
scattering rate for values $\Gamma = \alpha_g^2 M_0/2\delta M =  10^1,
10^2, \dots, 10^5$; this is the parameter that is most
directly associated with  the strength of the dark gauge
interaction since it multiplies the potential in (\ref{numeq}).  For each value of $\Gamma$, we explore a range of $\eta
= \alpha_g \mu /2 \delta M$ (recall that $\eta$ determines the range of
the interaction since $\mu$ is the gauge boson mass) that
includes the region where the rate of excitations is maximized.
For each value of $\Gamma$, we calculated contours of the positron
production rate $R_{e^+}$ in the $\eta$-$a$ plane.  However to make
the presentation  more concrete, instead of using
the dimensionless variable $a$, we momentarily assume that the model
should also account for the positron excesses seen by PAMELA and
Fermi/LAT, the latter of which suggests that $M_0\cong 1$ TeV
\cite{wimp-mass}.  This
enables us to plot the contours in the $\eta$-$\delta M$ plane.
An important question is whether the rate can ever be large enough
if $\delta M\ge 2 m_e$ since in the most natural models, each excited
state must have enough energy to decay into an $e^+$-$e^-$ pair.

In figure \ref{edmcont} we display results for three cases with the
BDM velocity dispersion (\ref{vbdm}) and different assumptions for
the density profile, and one example with the PDM dispersion
(\ref{vpdm}).  The largest rates are obtained in the former case,
using cuspy density profiles, and $\Gamma = 10^4$.  Even in this most
favorable case however, the mass splitting cannot exceed 600 keV if
the predicted rate is to match the observed one.  The cored BDM 
density profile and the PDM velocity profile give much smaller rates,
with  correspondingly smaller upper limits on $\delta M$, 170 keV and
200 keV respectively.

\subsection{Dependence on other parameters}

Ideally one would like to find examples where the predicted rate can
match observations for mass splittings $\delta M\gsim 2 m_e$, since
in the simplest models each excited DM state must decay into the 
ground state plus $e^+ e^-$.  Toward this end, we have explored the
dependence on other parameters which we previously held fixed.
For this purpose we fix instead the microscopic parameters at one
of the highest-rate examples found in fig.\ \ref{edmcont}.
Namely we take $\Gamma=10^4$, $\eta=50$, using the BDM
velocity profile, and Einasto density profile with $\alpha=0.145$.  
We vary the DM mass $M_0$,  the velocity dispersion parameter $v_0$,
and the circular velocity parameter $v_c$.  Recently ref.\
\cite{Reid} presented evidence, based on trigonometric parallaxes and
proper motions of masers in star-forming regions of the Milky Way,
 favoring a surprisingly large value
$v_c = 254\pm 16$ km/s.  The claim has been criticized in subsequent
references \cite{bovy,McMillan}; in particular ref.\ \cite{McMillan}
notes that the method used only constrains the ratio $v_c/R_0 \cong
30$ km/s/kpc, where $R_0$ is the distance to the galactic center.  We
therefore also consider the variation of $v_c$ and $R_0$ together.  

The results are shown in figure \ref{others}. The top left figure
shows that there is an optimal value of $M_0\cong 600$ GeV for
maximizing the rate, where $\delta M = 650$ keV gives the observed
rate.  This can be understood from the analytic approximation
(\ref{analytic}) where for fixed $\delta M$, the rate scales as $a^4
e^{-a}\sim M_0^{-4}e^{-c/M_0}$.  The bottom left figure shows  that
the dependence on $v_0$, which controls the overall size of the DM
velocity dispersion, is quite weak for values above the standard
one $v_0\cong 220$ km/s (see footnote \ref{f2} for explanation). 
As expected, decreasing $v_0$ below this value only reduces the rate. 

The right panels of figure \ref{others} show the dependence on  the
circular velocity $v_c$, which controls the escape velocity of the
DM.  Keeping $R_0$ fixed, the upper right panel demonstrates that
$v_c$ would have to be increased to  350 km/s to allow for $\delta M$
as large as $2 m_e$.  However increasing $R_0$ in a correlated way,
as suggested by ref.\ \cite{McMillan},  allows for a stronger effect,
because the cutoff $r_c$ on the radial integration in eq.\
(\ref{rate}) increases proportionally with $R_0$.  In this case, one
would need $v_c=280$ km/s and $R_0 = 9.4$ kpc.  Such a large value of
$v_c$ is $4\sigma$ away from the recent mean determination of $236\pm
11$ km/s \cite{bovy} and that of  $R_0$ is 2.5$\sigma$ and
3.1$\sigma$ outside the preferred values of refs.\
\cite{Ghez,Gillessen} respectively.

We conclude that, by taking all parameters and distribution functions
to their limits, it may be marginally possible to excite dark matter
with a mass splitting as large as $2m_e$, although most practitioners
would probably regard  the required values of $v_c=280$ km/s and
$R_0=9.4$ kpc as being unreasonably large.

\begin{figure}[t]
\smallskip \centerline{\epsfxsize=0.25\textwidth\epsfbox{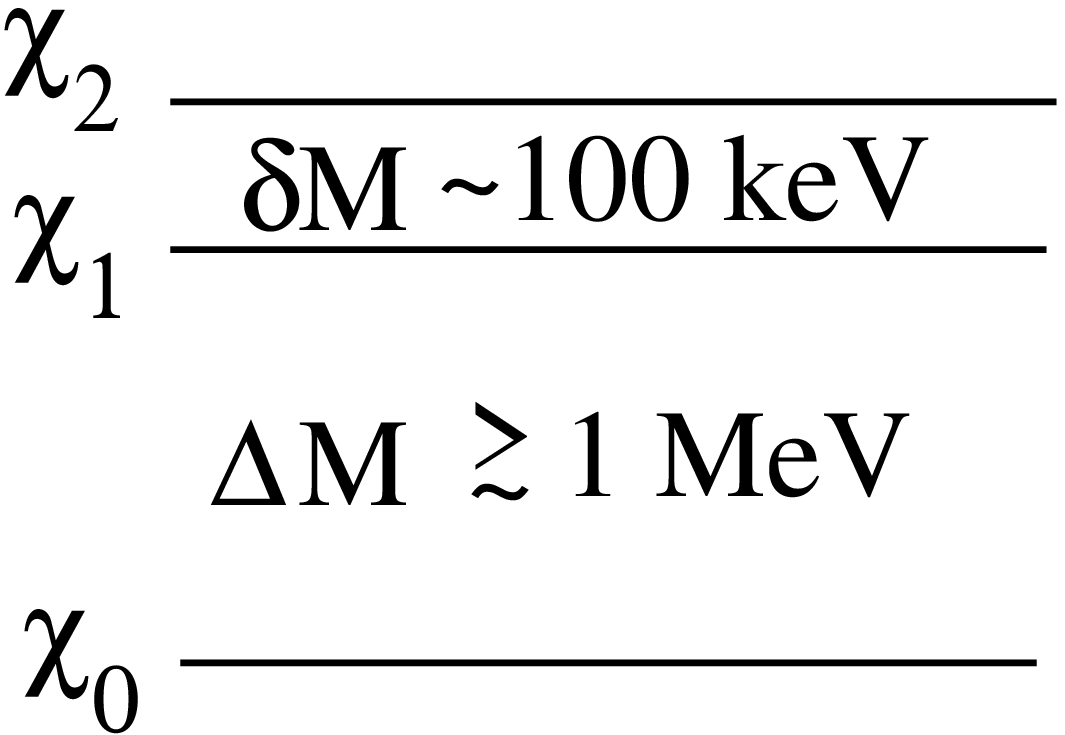}}
\caption{Spectrum of states for inverted mass hierarchy.}
\label{spect}
\end{figure}

\section{Alternative XDM scenarios and TeV dark matter}
\label{alternative}

In ref.\ \cite{CCF} it was pointed out that a long-lived species
$\chi_1$ with a mass splitting $\Delta M \sim 2 m_e$  above the
ground state $\chi_0$ could be excited to a third state $\chi_2$ with
splitting $\delta M \ll 2 m_e$ in an ``inverted mass hierarchy''
scenario, whose spectrum is shown in  fig.\ \ref{spect}.  Then the
XDM mechanism can work via the excitations
$\chi_1\chi_1\to\chi_2\chi_2$ followed by the decays $\chi_2\to\chi_0
e^+ e^-$.  The reduced value of $\delta M$ allows the rate to be
large enough to match observations without requiring extreme values
for parameters of the galactic velocity profile.   

Simple models of
nonabelian DM with hidden SU(2) gauge symmetry were constructed in
ref.\ \cite{CCF2} to realize this possibility. The scenario comes
with the expense of needing a late-time nonthermal origin for the DM,
as explained in \cite{CCF} (see also \cite{BPR}), since otherwise the
same process needed for positron production in the galaxy will
depopulate the $\chi_1$ states in the early universe.  If the DM
goes out of kinetic equilibrium early enough, a small remnant of its 
initial relic density can be maintained; ref.\ \cite{FSWY}
optimistically estimates that $\sim 1/10$ of the initially produced relic
density can survive.  In this case we would need parameters 
corresponding to a rate $R_{e^+}/R_{\rm obs} = 100$ to compensate
for the $\rho^2\sim 0.01$ suppression in the rate, relative to the
assumed case of a standard relic density.  The upper right-hand panel
of figure \ref{edmcont} (corresponding to BDM with a steep Einasto profile)
reveals such examples (the contours labeled ``2'') when $\Gamma
\sim 10^3-10^4$ and $\delta M \sim 50$ keV.

\begin{figure}[t]
\smallskip \centerline{\epsfxsize=0.4\textwidth\epsfbox{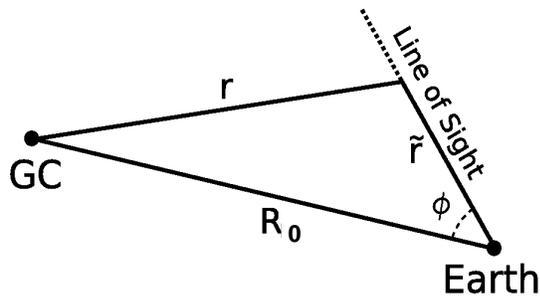}}
\caption{Relation between $r$, $\tilde r$ and $\phi$ for the line of
sight integral (\ref{Leq}).}
\label{geom}
\end{figure}

\begin{figure*}[t]
\smallskip \centerline{\epsfxsize=\textwidth\epsfbox{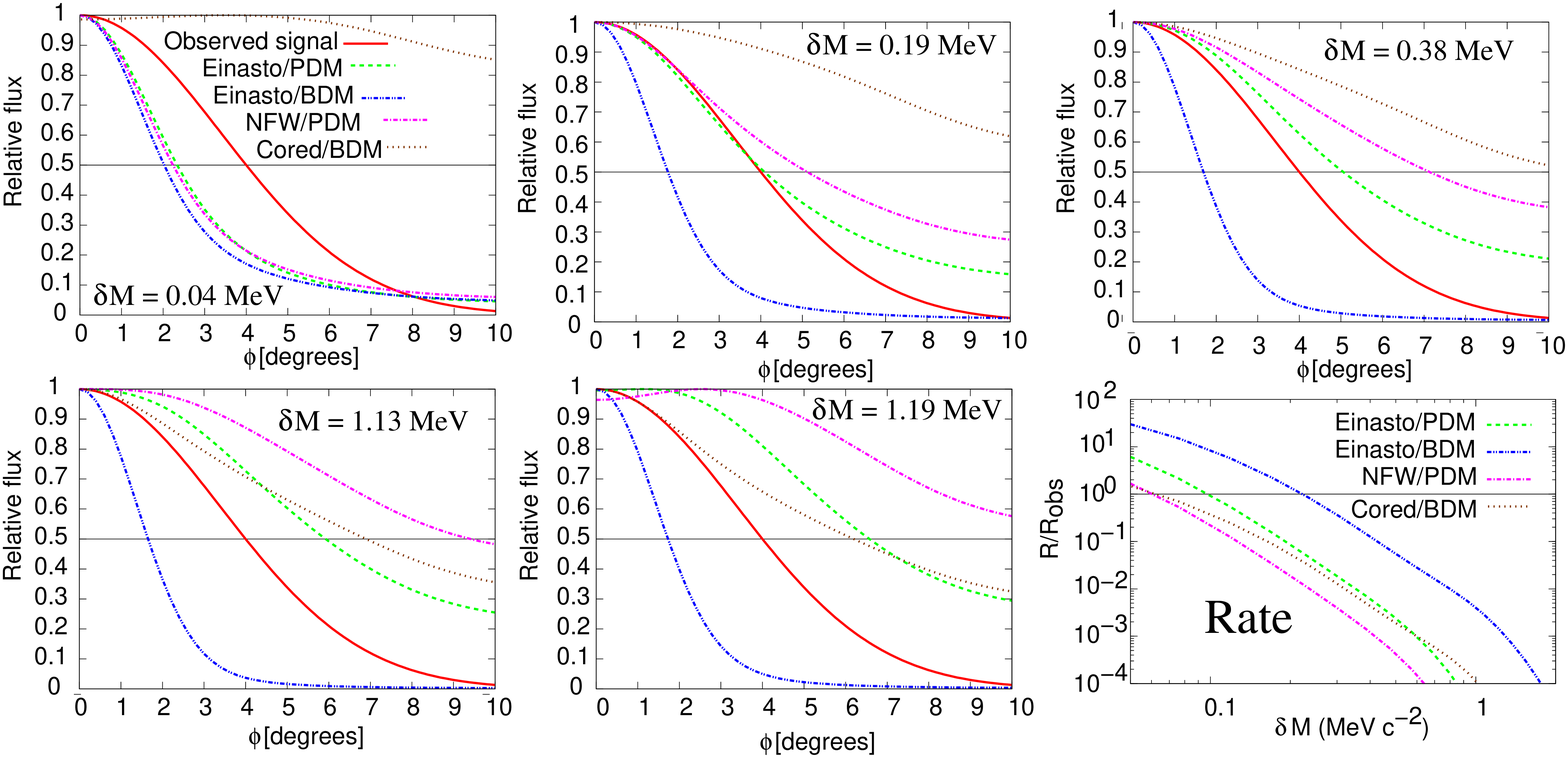}}
\caption{Comparison of the predicted and observed angular profiles
for $\Gamma = 100$, $\eta = 3.5$ and five values of $\delta M$, also varying
the density profiles.  All curves are normalized to 1 at $r=0$ to
emphasize the differences in shape.  
Bottom right graph shows dependence of the actual
rate on $\delta M$ for these same density profiles, assuming that
$M_0=1$ TeV.}
\label{shape}
\end{figure*}

In the case of nonthermally-produced intermediate states, 
in order for this component of the DM to
dominate over the conventionally produced thermal component of the
ground state, it is necessary to have an annihilation cross section 
that
is stronger than the one needed for the standard thermal relic
density, which was computed for SU(2) DM in ref.\ \cite{CCF2}.
This puts a constraint on the dark gauge coupling of triplet DM,
\beq
	\alpha_g > 0.03\, \left({M_0\over 1{\rm \ TeV}}\right)
\eeq
If we assume that $M_0\cong$ 1 TeV, and 200 MeV $<\mu< $ 1 GeV, the
preferred value for explaining excess leptons seen by the PAMELA and
Fermi/LAT detectors, this puts constraints on the dimensionless
parameters $\Gamma,\eta$:
\beqa
4.6 &<& \Gamma \times\left(\delta M \over  100\ {\rm MeV}\right) < 5100 \\
0.03 &<&\eta \times\left(\delta M \over  100\ {\rm MeV}\right) < 5
\eeqa
where we have also imposed that $\alpha_g < 1$.
Referring to fig.\ \ref{edmcont}, we see that these can be satisfied
for the top panels, which use Einasto profile and BDM velocity
dispersion, even at $\Gamma = 10$.  There thus seems to be
considerable room in the parameter space for the XDM explanation of
low-energy galactic center positrons, if one accepts the inverted
mass hierarchy hypothesis and a relic density of intermediate mass
states.

\section{Angular distributions}
\label{angular}

Further constraints can be obtained by computing the expected angular
profiles of the produced positrons.  This gives one-sided bounds,
because there is considerable uncertainty in the distance that a
positron could propagate away from the galactic center before
annihilating.  But a model that predicts too wide an angular
profile can be excluded since propagation effects will never make
the true profile more narrow. 

Neglecting propagation effects, the angular distribution of positrons
is given by an integral similar to that in (\ref{rate}), but instead
of integrating over the volume, one must integrate along the 
line of sight.  We define a luminosity $L(\phi)$ by

\beq
L(\phi) = \int_0^{\infty}  
\VEV{\sigma v_{\rm rel}}(r)\, n^2(r)\, \ud \tilde r\,
\label{Leq}
\eeq
where $\tilde r$ is distance along the line of sight, and 
$r^2(\tilde r,\phi)= R_0^2 + \tilde r^2 -2 \tilde r R_0\cos\phi$,
as illustrated in fig.\ \ref{geom}.  The observed profile is broadened
by the SPI instrumental resolution of $3^\circ$ full width at half
maximum, so we smooth it using a gaussian, 
\beq
	L_{\rm obs}(\phi) = N(c)\int d\phi_0\, e^{-c(\phi-\phi_0)^2)}
	L(\phi_0)
\eeq
where $c = \pi(3^\circ /180^\circ)/(2^{3/2}\ln 2)$.

Our results for the angular profile are shown in fig.\ \ref{shape},
for sample microphysics parameters $\Gamma=100$ and $\eta=3.5$.
We find that the shapes are relatively insensitive to these values,
and rather depend more strongly on the DM density and velocity
profiles, and the parameters $a,b$ defined in (\ref{abdef}).
For the first five panels of this figure, we have ignored the overall
rate and normalized all curves to match the observed profile at the
galactic center, to better visualize the differences in widths of
the predicted signals.  The bottom right panel shows the real rate
corresponding to each assumed density profile, as a function of 
$\delta M$ (assuming $M_0=1$ TeV as before).  

One can see that for the larger values of $\delta M$,  the pure dark
matter velocity profiles (as well as cored BDM) produce  too wide a profile to match the
observations, while the cuspy BDM case gives a narrower result that can be
consistent if the positrons  travel a distance of order 1 kpc before
annihilating, as has been argued is possible \cite{transport}.
  This indicates that the kinematic advantage of having
higher central velocities in the BDM scenario is more important than
the boost to the cross section that one would expect from the
Sommerfeld enhancement at lower velocities.  Thus not only are the 
cuspy BDM profiles more easily able to match the observed rate, but
they are also more consistent with the observed angular distribution.

\section{Conclusion}
\label{conclusion}

In this paper we have tried to give the most quantitative treatment
to date of the excited dark matter mechanism for producing positrons
at the galactic center.  A main technical improvement was the
numerical computation of the excitation scattering cross section,
which is nonperturbative and guaranteed to satisfy unitarity
constraints.  A second difference relative to previous treatments is
that we included the effects of baryons on the velocity dispersion of
DM in the inner kpc of the galaxy, which can significantly boost the
rate of excitations.  In addition we considered NFW and Einasto
profiles for the DM density, varying parameters determining the
cuspiness of the distributions, as well as those affecting the
velocity distributions.  

Even making all the most optimistic assumptions for increasing the
rate of DM excitations, we were not able to find realistic parameter
values which yield a large enough rate, if the mass gap between the
ground state $\chi_0$ and excited state $\chi_1$ of the DM is
sufficient for  the subsequent decay $\chi_1\to \chi_0\, e^+ e^-$
to produce the positron.  The largest mass gap we found
consistent with the observed rate was $\delta M \cong 650$ keV,
assuming the standard values $v_c = 220$ km/s  and $R_0 = 8$ kpc,
respectively, for the circular velocity of the sun around and its
distance to the galactic center.  By pushing these values to  $v_c =
280$ km/s and $R_0 = 9.4$ kpc, we can barely accommodate a mass gap
of $\delta M = 2 m_e$, but these choices are respectively $4\sigma$
and $3\sigma$ away from the mean values.

One way to ease the tension would be to consider models where the
excited states are charged and thus able to decay into single
electrons or positrons; this would lower the required mass gap to
just $\delta M = m_e$.  But this requires a great deal more
model-building gymnastics than the case of a neutral excited state. A
theoretically more appealing alternative is the case of three DM states where
$\chi_0\chi_0\to \chi_1\chi_2$, in which the mass gap $\delta M_{01}$
between $\chi_0$ and $\chi_1$ is much smaller than $\delta M_{02} > 2
m_e$ \cite{Cheung,CCF2}.  For example if $\delta M_{01}=0$, then the effective  $\delta
M$ which determines the threshold velocity is half as large as in the
models that produce two of the same excited state.  This possibility
can arise if the hidden sector gauge group is  SU(2)$\times$U(1) and
gets completely broken, for example.  However to see if this class of
models can really have a larger rate would require solving the
Schr\"odinger equation in the three-state system, so while it is
plausible that such asymmetric excitations could increase the rate,
it is not proven by our analysis.

Another way of getting around our no-go result is to assume the
existence of a stable or metastable population of already excited
states $\chi_1$, which need only be further excited to $\chi_2$
through a small mass gap $\delta M_{12}$, if $\delta M_{02}$ is
assumed to be greater than $2 m_e$ so that the decay $\chi_2\to
\chi_0\,
e^+e^-$ is allowed.  We refer to this as the inverted mass hierarchy
scenario.  This idea comes with new complications, since it is
difficult to prevent the excitation process from occuring earlier in
the history of the universe, to maintain the population of $\chi_1$
to the present day.  However there is an  intriguing possibility to
overcome this by assuming the present DM particles are products of
the late decay of a much heavier predecessor, and thus were initially
relativistic, at a time when they would normally have been
nonrelativistic had they been produced through conventional
freeze-out \cite{CCF}.  The higher velocity suppresses the Sommerfeld
enhancement at early times, when it would have the undesirable effect
of depleting the metastable intermediate DM states.  We hope to
examine this scenario more carefully in future work.  

We have not tried to impose in detail the additional
constraints on the model which arise if one would like it to also
account for the PAMELA and Fermi/LAT excess electrons/positrons,
although we did focus on  TeV scale DM for that purpose.  If the
DM explanation of the high-energy leptons is not ruled out by 
upcoming analyses, this would be an interesting next step.   It might
also be worthwhile to investigate the effect of a nonspherically
symmetric DM halo \cite{jmr} on the rate of positron production.
 
\centerline{$\phantom{.}$}
{\bf Acknowledgment:}  We thank Gil Holder for valuable discussions
about the galactic parameters.

\appendix

\section{Born approximation cross section}
\label{born}

For comparison with the nonrelativistic excitation cross section 
from numerical solution of the Schr\"odinger equation, we here
give the quantum field theoretic expression in the Born approximation.
The spin-averaged squared matrix element is
\begin{widetext} 
\begin{eqnarray}
|\mathcal{M}|^2=4g^4\left\{\frac{4}{(t-\mu^2)^2}\left[2(s^2+u^2)+4M_+^2t-M_+^4+M_-^4-6M_+^2M_-^2\right]\right.\nonumber\\
                  +\frac{4}{(u-\mu^2)^2}\left[2(s^2+t^2)+4M_+^2u-M_+^4+M_-^4-6M_+^2M_-^2\right]\nonumber\\
                    \left.  +\frac{4}{(t-\mu^2)(u-\mu^2)}\left[4s^2-8M_+^2s+2M_-^2s+3(M_+^2-M_-^2)^2\right] \right\}
\end{eqnarray}
where $g$ is the couping constant, $s$, $t$, $u$ are the 
Mandelstam variables, $M_\pm =M_1\pm M_0$,  and 
$M_1$, $M_0$ and $\mu$ are the dark matter 
and gauge boson mass respectively.

In the center-of-mass frame,
\begin{eqnarray}
s=E_{cm}^2, \, t=-2p^2(1-\cos(\theta)), \, u=-2p^2(1+\cos(\theta))
\end{eqnarray}
so the cross section is
\begin{eqnarray}
\sigma &=&\frac{1}{E_{cm}^2v_{rel}}\int\frac{d\Omega_{cm}}{4\pi}\frac{1}{8\pi}\frac{2|p|}{E_{cm}}|\mathcal{M}|^2\nonumber\\
              &=&g^4\frac{|p'|}{2\pi E_{cm}^3v_{rel}}\left\{-\frac{8s^2-16M_+^2s+4M_-^2s+6(M_+^2-M_-^2)^2}{pp'(\mu^2+p^2+p'^2)}\ln(\frac{\mu^2+(p+p')^2}{\mu^2+(p-p')^2})\right.\nonumber\\
                &-&\frac{64\mu^4+192\mu^2p^2+192\mu^2p'^2+64\mu^2M_+^2+160p^4+160p'^4+192p^2p'^2+32s^2-16M_+^4+16M_-^4-96M_+^2M_-^2}{(\mu^2+(p-p')^2)(\mu^2+(p+p')^2)}\nonumber\\
                 &+&\left.\frac{16(2p^2+2p'^2+M_+^2+\mu^2)}{pp'}\ln(\frac{\mu^2+(p+p')^2}{\mu^2+(p-p')})\right\}
\end{eqnarray}

We rewrite the kinematic variables using the dimensionless
quantity $\Delta$, 
\begin{eqnarray}
v_{rel}=2v_t\sqrt{\Delta}, \,\, p=M_0v_t\sqrt{\Delta}, \,\,
p'=M_0v_t\sqrt{\Delta -1}\nonumber\\
 E_{cm}=2M_0\sqrt{1+v_t^2\Delta }, \,\,
M_+=2M_0\left(1+\frac{v_t^2}{4}\right), \,\, M_-=\frac{M_0}{2}v_t^2
\end{eqnarray}
Using $v_t\ll 1$ and $\delta M \ll \mu\ll M_0$, we obtain the 
approximate expression (\ref{sigma_born}) for the cross section.

\end{widetext}

\section{Escape velocity}
\label{vesc_app}

To be self-contained, we reproduce here the argument of ref.\
\cite{xdm} for the $r$-dependence of the escape velocity.  It is
important to notice that dark matter does not dominate the mass of
the inner galaxy.  To account for the gravitational effect of the
baryons, one can use the fact that the circular velocity 
$v_c$ (the velocity of a test mass on a circular orbit around the
galactic centre), is nearly constant out to radii or order $r_{-2}$.
We make the simplifying assumptions that there is no mass beyond 
$r=r_{-2}$ and that the density is spherically symmetric.

The gravitational
potential is 
\beq
\Phi = -G\int_r^{\infty} {M(r')\over r'^2} \ud r'.
\eeq
where $M(r)$, the mass within radius $r$, can be inferred from $F=ma$
for a test particle of mass $m$: $mv_c^2/r = GmM(r)/r^2$.  Hence
$M(r) = r v_c^2/G$ for $r< r_{-2}$ and $M=r_{-2} v_c^2/G$ for $r > 
r_{-2}$.  The integral for $\Phi$ becomes
\beq
\Phi =  v_c^2\left\{\begin{array}{ll}
\ln\left(\frac{r}{r_{-2}}\right)-1,& r<r_{-2}\\
-\frac{r_{-2}}{r},& r<r_{-2}\end{array}\right.
\eeq
The escape velocity is given by $\frac12 v_{\rm esc}^2 + \Phi=0$,
leading to eq.\ (\ref{vesc_eq}).

\end{document}